\documentstyle[aps,psfig,amsmath]{revtex}
\newcommand{\be}{\begin{equation}}
\newcommand{\ee}{\end{equation}}
\newcommand{\bea}{\begin{eqnarray}}
\newcommand{\eea}{\end{eqnarray}}
\newcommand{\bef}{\begin{figure}}
\newcommand{\eef}{\end{figure}}
\newcommand{\bm}{\bibitem}

\newcommand{\bet}{\beta}
\newcommand{\th}{\theta}
\newcommand{\lm}{\lambda}
\newcommand{\sg}{\sigma}
\newcommand{\de}{\delta}
\newcommand{\De}{\Delta}
\newcommand{\qls}{q_\lm q_\sg}
\newcommand{\gf}{\gamma_5}
\newcommand{\ep}{\epsilon}
\newcommand{\gamu}{\gamma_{\mu}}
\newcommand{\gamU}{\gamma^{\mu}}
\newcommand{\om}{\omega}
\newcommand{\rw}{\rightarrow}

\newcommand{\cl}{{\cal{L}}}
\newcommand{\co}{{\cal{O}}}
\newcommand{\bp}{{\boldsymbol{\pi}}}
\newcommand{\bt}{{\boldsymbol{\tau}}}
\newcommand{\ba}{{\boldsymbol{a}}}
\newcommand{\F}{F_\pi}
\newcommand{\dmd}{\partial_\mu}
\newcommand{\dmu}{\partial^\mu}
\newcommand{\vpamu}{v_\mu+a_\mu}
\newcommand{\vmamu}{v_\mu-a_\mu}

\newcommand{\omk}{\omega_k}
\newcommand{\ov}{\overline}
\newcommand{\bq}{\ov{q}}
\newcommand{\bu}{\ov{u}}
\newcommand{\vk}{\vec{k}}

\newcommand{\la}{\langle}
\newcommand{\ra}{\rangle}

\newcommand{\kp}{k^\prime}
\newcommand{\pp}{p^\prime}
\newcommand{\ip}{{i^\prime}}
\newcommand{\spr}{s^\prime}
\newcommand{\dab}{\delta^{i\ip}}
\newcommand{\gm}{\gamma}
\newcommand{\ta}{\tau^i}
\newcommand{\qs}{q\!\!\!/}
\newcommand{\ps}{p\!\!\!/}
\begin{document}

\setcounter{page}{1}

\title{Pion parameters in nuclear medium from chiral perturbation theory 
       and virial expansion}

\author{S. Mallik} 
\address{Saha Institute of Nuclear Physics,
1/AF, Bidhannagar, Kolkata-700064, India} 

\author{Sourav Sarkar} 
\address{Variable Energy Cyclotron Centre, 
1/AF, Bidhannagar, Kolkata-700064, India}


\maketitle

\begin{abstract} 
We consider two methods to find the effective parameters of the pion 
traversing a nuclear medium. One is the first order chiral perturbation
theoretic evaluation of the pion pole contribution to the two-point function 
of the axial-vector current. The other is the exact, first order virial 
expansion of the pion self-energy. We find that, although the results of
chiral perturbation theory are not valid at normal nuclear density, those
from the virial expansion may be reliable at such density. The latter
predicts both the mass-shift and the in-medium decay width of the pion to be
small, of about a few MeV.
\end{abstract}


\section{Introduction}

A considerable amount of work at finite temperature and chemical potential has 
been devoted to determining the effective parameters of strongly interacting 
particles in different media~\cite{Gasser1,Goity,Schenk,Smilga,Mallik1,Rapp}. 
The results obtained are useful not only in analyzing the heavy ion collision 
experiments and properties of the early universe at different epochs, but 
also in extracting indications of an eventual phase transition.

The case of pion appears to be the simplest to study. Being the Goldstone boson
of the spontaneously broken chiral symmetry of QCD, its interactions with 
itself and other particles are highly restricted by this symmetry, leading 
to the effective theory of QCD, called chiral perturbation theory 
($\chi PT$)~\cite{Weinberg1,Gasser2}. At finite temperature, one has the 
further advantage of having again only pions dominating the heat bath. Thus 
$\chi PT$ provides a reliable method to calculate the pion parameters at 
finite temperature ~\cite{Gasser1,Goity,Schenk}.

It is natural to apply $\chi PT$ to calculate the pion parameters in nuclear
medium~\cite{Thorsson,Meissner}. Although the method parallels that followed 
for the case in a heat bath, the results calculated here to leading order 
may have restricted validity, due to the presence of baryonic resonances 
close to the $\pi N$ threshold. Similar difficulties also appear in 
determining the nucleon parameters at finite temperature~\cite{Smilga}.

In this work we compare the $\chi PT$ result with that of the (first order) 
virial expansion of the pion self-energy \cite{Smilga,Jeon,Mallik2}. The latter  
gives the shifted pole position in terms of an integral over the product of 
the density distribution function times the $\pi N$ scattering amplitude 
obtainable from experiment. It is thus free from the difficulty encountered 
in calculating the amplitude and is valid as long as the nucleon `gas' is 
dilute enough.

The virial formula has also been employed earlier to the same problem, but only
in an approximated version \cite{Ericson,Migdal,Waas}. If, however, the 
amplitude varies appreciably in the range of integration, in particular, 
if it changes sign --  as is the case here -- this version is not justified.

We first rederive the results of $\chi PT$ by evaluating the axial-vector
current correlation function to one loop, using the in-medium Feynman rules
for the original chiral Lagrangian in presence of external fields. Besides
completeness, its purpose is to show that this conventional framework is
quite simple, without requiring functional integration over the nucleon
field to produce a `new' effective Lagrangian~\cite{Meissner}. We then 
derive the exact, first order virial expansion for the pion self-energy 
and evaluate it with experimental data.    
   
Sec.~II reviews briefly $\chi PT$, constructing the Lagrangian for the 
$\pi N$ system ~\cite{Gasser3,Fettes}. In sec.~III we work 
out the shift in the mass and the decay constant of the pion using this 
Lagrangian. Next we derive the virial formula and evaluate the pole shift 
in sec.~IV. In sec.~V we discuss the limitations of these methods.

\section{Chiral Perturbation Theory}
\setcounter{equation}{0}
\renewcommand{\theequation}{2.\arabic{equation}}

The Lagrangian of QCD with two massless quark flavors is
\be
{\cal L}^{(0)}_{QCD}=i\ov q\gamU\dmd q + \cdot\cdot\cdot\cdot,
~~~~~~~q=\left(\begin{array}{c}u\\d\end{array}\right)~~,
\ee
where the dots denote terms involving other fields. If we split the quark
field into its right and left handed parts, $q_{R,L}= \frac{1}{2} 
(1\pm\gamma_5)q$, it is clear that ${\cal L}^{(0)}_{QCD}$ is invariant under
the symmetry group $G=SU(2)_R\times SU(2)_L$ of independent, global $SU(2)$
transformations on $q_R$ and $q_L$,
\be
q_R\rw g_R q_R~~,~~q_L\rw g_L q_L~~,~~~~~g_{R,L}\in SU(2)_{R,L}~.
\ee
Phenomenology suggests strongly that the symmetry of the Lagrangian is broken
spontaneously by the vacuum state to the diagonal subgroup $H=SU(2)_V$,
giving rise to the pionic degrees of freedom.

In $\chi PT$ one derives the transformation rules for the observed 
Goldstone and non-Goldstone fields from the above symmetry of the 
underlying QCD theory. It turns out that the Goldstone fields 
$\pi^i(x),~~i=1,2,3$ are collected in the form of a unitary matrix
\be
u(x) = e^{i\pi (x)/{2\F}}~,~~~~~~~~\pi (x)=\sum_{i=1}^3\pi^i(x)\ta~~,
\ee
where the constant $\F$ can be identified with the pion decay constant, 
$\F =92.4$ MeV and $\ta$ are the Pauli matrices. Then the matrix $u$ 
transforms under $G$ according to 
\be
u\rw g_R u h^\dag=hug_L^\dag~~,
\ee
where the group element $h(\pi)\in SU(2)_{V}$. Notice that $h$ is $x$-dependent 
due to its dependence on $\pi^i(x)$. However the square of this matrix
$u^2=U$ has the global transformation rule
\be
U\rw g_R\,U\,g_L^\dag.
\ee
On the other hand, the non-Goldstone, nucleon doublet field $\psi(x)$
transforms as 
\be
\psi\rw h\psi~~,
~~~~~~~\psi=\left(\begin{array}{c}p\\n\end{array}\right)~~.
\ee 

There are two Noether currents following from the symmetry of 
${\cal L}^{(0)}_{QCD}$, namely the vector and axial vector currents,
\be
V_\mu^i(x)=\bq(x)\gamu\frac{\ta}{2}q(x),~~~~~~~~~
A_\mu^i(x)=\bq(x)\gamu\gamma_5\frac{\ta}{2}q(x)~.
\ee
The evaluation of the correlation functions of the currents is 
most conveniently carried out in the external field method~\cite{Gasser2}. 
Although we are interested here in such a function of the axial-vector
current only, we couple both the currents to the external fields 
$v_\mu^i(x)$ and
$a_\mu^i(x)$ to reveal the full symmetry of the underlying theory. Thus the
original Lagrangian extends to
\bea
&&{\cal L}^{(0)}_{QCD}+v_\mu^i(x)V^\mu_i(x)+a_\mu^i(x)A^\mu_i(x)\nonumber\\
&=&i\,\bq_R\gamU\{\dmd-i(\vpamu)\}q_R+
i\,\bq_L\gamU\{\dmd-i(\vmamu)\}q_L+\cdot\cdot\cdot
\eea
where $v_\mu(x)$ and $a_\mu(x)$ are the matrix valued external vector and
axial vector fields,
\be
v_\mu(x)=\sum_{i=1}^3 v_\mu^i(x)\frac{\ta}{2}~~~,~~~~~~
a_\mu(x)=\sum_{i=1}^3 a_\mu^i(x)\frac{\ta}{2}.
\ee
The extended Lagrangian is now invariant locally, i.e. under $x$-dependent
symmetry transformations on $q_R$ and $q_L$, if the external vector and  
axial-vector fields are also subjected to the appropriate gauge transformations.

The presence of the external fields in the underlying theory
and the associated gauge invariance can be readily incorporated in the
effective theory. All we have to do is to replace the ordinary
derivatives by the covariant derivatives~\cite{Comment1}. Thus we have for $U$,
\be
D_\mu U=\dmd U-i(\vpamu)U+iU(\vmamu),
\ee
and for $\psi$,
\bea
{\cal D}_\mu\psi&=&\dmd\psi+\Gamma_\mu\psi~~,\nonumber\\
\Gamma_\mu&=&\frac{1}{2}\left(u^\dag[\dmd-i(\vpamu)]u+u[\dmd-i(\vmamu)]u^\dag
\frac{}{}\right)~~.
\eea
The building elements for the effective Lagrangian at this stage are thus
$U$, $D_\mu U$, $\psi$ and ${\cal D}_\mu\psi$.

A simplification in the construction of the effective Lagrangian emerges by
noting that the variables ($U$, $D_\mu U$) transform under the full group $G$,
while ($\psi$, ${\cal D}_\mu\psi$) transform only under the 
unbroken subgroup $H$.
We may take advantage of the mixed transformation property of $u$ to redefine
the former type of variables so as to transform under $H$ only. Thus one 
introduces the variable~\cite{Gasser3},
\be
u_\mu=iu^\dag D_\mu U u^\dag =u_\mu^\dag~~,
\ee
replacing $U$ and $D_\mu U$ \cite{Comment1}. Any term in the Lagrangian 
that is constructed 
out of these variables so as to be invariant under $H$ will also be 
automatically invariant under $G$. 

We now write the effective Lagrangian of $\chi PT$ for the $\pi N$ system as
\[\cl_{ef\!f} = \cl_\pi +\cl_N, \]
where $\cl_\pi$ is the well-known pion Lagrangian~\cite{Gasser2}, which to 
leading order is given by
\be
\cl_\pi=\frac{\F^2}{4}\{\la D_\mu U D^\mu U^\dag\ra+m_\pi^2\la U+U^\dag \ra\}~,
\ee
$\la\cdot\cdot\cdot\ra$ denoting trace over the 2$\times$2 isospin matrices.
The pieces in $\cl_N$ to first and second order,
\be
\cl_N=\cl_N^{(1)}+\cl_N^{(2)}~,
\ee
are
\be
\cl_N^{(1)}=\ov \psi(i{\cal D}\!\!\!\!/-m_N)\psi+\frac{g_A}{2}
\ov\psi u\!\!\!/\gf\psi~,
\ee
and
\be
\cl_N^{(2)}=c_1m_\pi^2\la U+U^\dag \ra\ov\psi\psi - \frac{c_2}{4m_N^2}
\la u_\mu u_\nu\ra(\ov \psi{\cal D}^\mu{\cal D}^\nu\psi + h.c.)
+\frac{c_3}{2}\la u_\mu u^\mu\ra\ov\psi\psi-\frac{c_4}{4}\ov\psi\gm^\mu\gm^\nu
[u_\mu,u_\nu]\psi~.
\ee
We shall use vertices in $\cl_N^{(1)}$ to second order
and those in $\cl_N^{(2)}$ to first order in our perturbative calculations.
Here $g_A$ turns out to be the axial-vector coupling constant appearing in the
neutron beta decay, $g_A$=1.27. The coupling constants $c_1,\,c_2$ and $c_3$ 
are determined from the experimental data for $\pi N$ scattering in the low
energy region and its extrapolation inside the Mandelstam triangle, where it
is compared with the $\chi PT$ evaluation, getting~\cite{Buettiker,Fettes},
\[c_1=-0.81\pm 0.12\,\,{\rm GeV}^{-1},~~c_2=3.2\pm 0.25\,\,{\rm GeV}^{-1}, 
~~c_3=-4.66\pm 0.36 \,\,{\rm GeV}^{-1}~. \]

Expanding out in the pion field and setting $v_\mu$=0, we bring out 
explicitly the vertices, contributing to the pion pole diagrams to one 
loop~\cite{Comment2},
\bea
\cl_\pi&=&\cl_\pi^{(0)}-\F\dmd\bp\cdot\ba^\mu~,\nonumber\\
\cl_N^{(1)}&=&\cl_N^{(0)}-\frac{g_A}{2\F}\ov\psi\gamU\gf\dmd\bp\cdot\bt\psi
+\frac{g_A}{2}\ov\psi\gamU\gf\ba_\mu\cdot\bt\psi~,\nonumber\\
\cl_N^{(2)}&=&-\frac{2m_\pi^2c_1}{\F^2}\bp\cdot\bp\,\ov\psi\psi
-\frac{c_2}{m_N^2}\left(\frac{1}{\F^2}\dmd\bp\cdot\partial_\nu\bp
-\frac{2}{\F}\dmd\bp\cdot\ba_\nu\right)\ov\psi\dmu\partial^\nu\psi\nonumber\\
&&+c_3\left(\frac{1}{\F^2}\dmd\bp\cdot\dmu\bp
-\frac{2}{\F}\dmd\bp\cdot\ba_\mu\right)\ov\psi\psi~.
\label{lag_int}
\eea
where the superscript `0' indicates free Lagrangian densities.

\section{Mass and coupling shifts}
\setcounter{equation}{0}
\renewcommand{\theequation}{3.\arabic{equation}}

The two point correlation function of the axial vector current in a medium is 
given by the ensemble average,
\be
i\int d^4x\, e^{iq\cdot x}\,Tr\,[e^{-\bet (H-\mu N)}A_\lm^i(x)\,A_\sg^\ip(0)]
/Tr\,[e^{-\bet (H-\mu N)}]~.
\label{ensavg}
\ee
Here $H$ is the QCD Hamiltonian, $\bet$ is the inverse temperature and
$N$ is the number operator for nucleons with chemical potential $\mu$. Our
aim is to find the corrections to the pion pole in this correlation function 
due to interaction of pion with nucleons in medium at zero temperature.

We shall work in the real time formulation of the field theory in medium 
\cite{Niemi}. Here
the perturbation expansion proceeds as in the conventional (vacuum) field
theory, except that the propagators assume the form of 2$\times$2 matrices. For
the pole term to one loop, it however suffices to work as in the vacuum field 
theory, only replacing the vacuum propagators by the 11-component of the
corresponding propagators in medium.

So we first calculate the vacuum correlation function
\be
i\int d^4x\, e^{iq\cdot x}\,\la 0|\,T\,A_\lm^i(x)\,A_\sg^\ip(0)|0\ra~,
\label{2point_vac}
\ee
in $\chi PT$. We recall that the generating functional of QCD,
\be
\la 0|\,T\,e^{i\int\,d^4x\,a^i_\mu(x)A^\mu_i(x)}|0\ra~,
\label{genfuncQCD}
\ee
is represented in $\chi PT$ by
\be
\la 0|\,T\,e^{i\int\,d^4x\,\cl_{int}(\pi,\psi,a)}|0\ra~,
\label{genfunceff}
\ee
where $\cl_{int}$ is obtained from Eqs.(2.17).
Since the two point function (\ref{2point_vac}) is the coefficient of the
term quadratic in $a_\mu(x)$ in the expansion of the generating functional
(\ref{genfuncQCD}), we find these quadratic terms from the functional
(\ref{genfunceff}) of the effective theory.

\bef[h]
\centerline{\psfig{figure=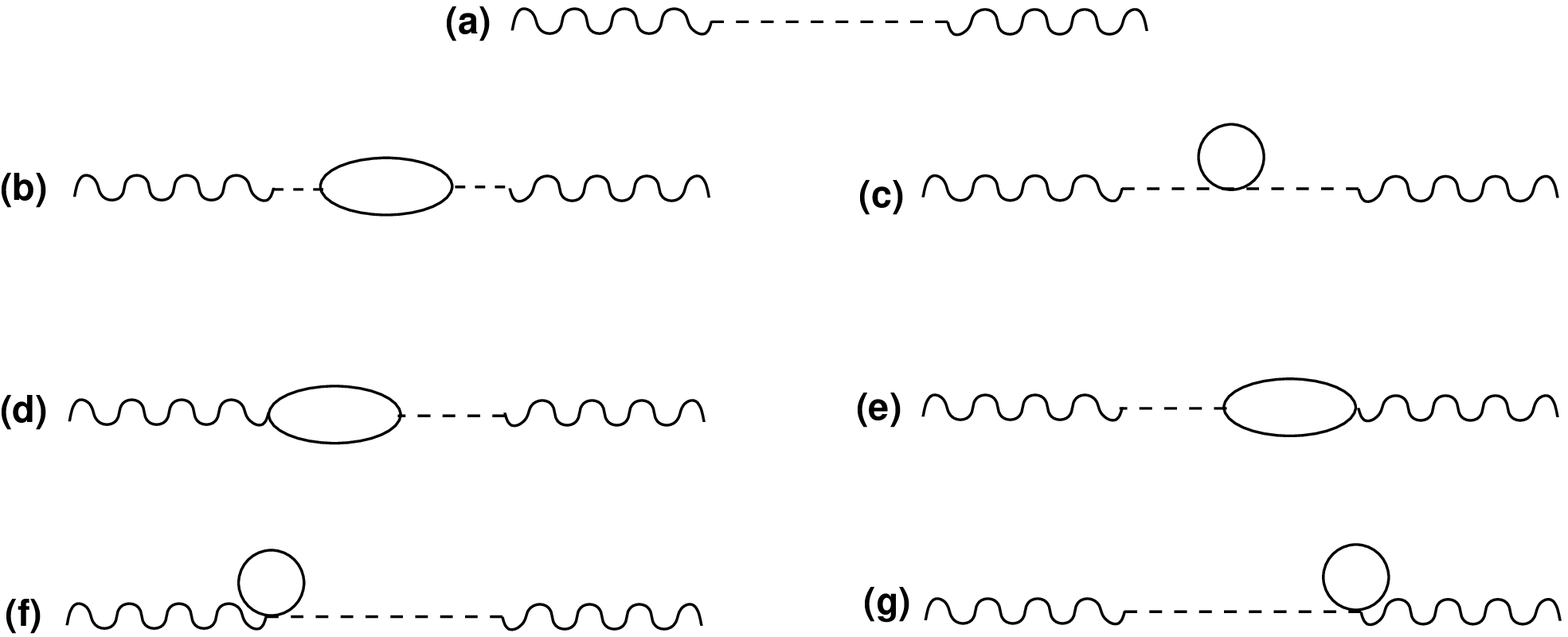,height=6.5cm,width=9cm}}
\caption{Feynman diagrams for the two point function to one loop. Only
loops with nucleons are considered.}
\eef

Fig.~1 shows the free pion pole diagram (a) giving the amplitude,
\be
\dab\qls i\F^2\De(q)~,~~~~~~\De(q)=i/(q^2-m_\pi^2+i\ep)~,
\label{freepole}
\ee
together with all one loop corrections to it, relevant in the
nuclear medium. Let us calculate the self-energy diagram (b) to
illustrate the method. Its contribution may be written in the form
\be
\dab\qls i\F^2\De(q)\{-i\Pi(q)\}\De(q)~,
\ee
which modifies the free pion pole term (\ref{freepole}) to
\be
\dab\qls\frac{-\F^2}{q^2-m_\pi^2-\Pi(q)}~.
\label{modpole}
\ee

To calculate the self-energy function $\Pi(q)$ of the pion in the
nuclear medium, we first write it in vacuum,
\be
\Pi^{(0)}(q)=\frac{ig_A^2}{2\F^2}\int\frac{d^4p}{(2\pi)^4}
tr\left[\qs\gf(\ps+m_N)\qs\gf(\ps-\qs+m_N)\right]\De^N(p)\De^N(p-q)~,
\label{se_vac}
\ee
where $\De^N(p)$ is the vacuum nucleon propagator after extracting the
factor $(\ps+m_N)$, $\De^N(p)=i/(p^2-m_N^2+i\ep)~.$
Following our discussion above, we now replace the vacuum nucleon propagator
in Eq.~(\ref{se_vac}) by its 11-component in nuclear medium
$(E_p=\sqrt{\vec{p}~^2 +m_N^2})$,
\be
\De^N_{11}(p)=\frac{i}{p^2-m_N^2+i\ep}-2\pi\left\{n^+(E_p)\th(p_0)
+n^-(E_p)\th(-p_0)\right\}\de(p^2-m_N^2)
\label{nprop11}
\ee
where $n^{\pm}(E_p)$ are the distribution functions respectively for the
nucleon and the antinucleon,
\be
n^{\pm}(E_p)=\frac{1}{e^{\bet(E_p\mp\mu)}+1}~.
\ee
As the temperature goes to zero, we get for $\mu > 0$,
\be
n^+(E_p)\rw\th(\mu-E_p)~,~~~n^-(E_p)\rw 0~.
\ee
The density dependent part of the self-energy in the medium is then obtained as
\be
\Pi^{(n)}(q)=-8g_A^2\frac{m_N^2}{\F^2}q^4\int\frac{d^4p}{(2\pi)^3}
\,\,\frac{\de(p^2-m_N^2)\,\th(\mu-p_0)\,\th(p_0)}{-4(p\cdot q)^2+q^4}~.
\ee
Consider the pion to be at rest $(\vec{q}=0)$ in the medium, 
when it simplifies to
\bea
\Pi^{(n)}(q_0,\vec q=0)&=&-8g_A^2\frac{m_N^2}{\F^2}q_0^2\int\frac{d^3p}
{(2\pi)^3 2E_p}\,\,
\frac{\th(p_F-|\vec p|)}{-4E_p^2+q_0^2}\nonumber\\
&=&\frac{g_A^2\ov n}{4\F^2m_N}\,q_0^2~,
\label{se_corr}
\eea
where $\ov{n}$ is the nucleon number density in symmetric nuclear matter,
\be
\ov n=4\int\frac{d^3p}{(2\pi)^3}\,\th(p_F-|\vec p|)=\frac{2p_F^3}{3\pi^2}~,
\ee
$p_F$ being the Fermi momentum, $p_F=\sqrt{\mu^2-m_N^2}$.
Assuming that the vacuum part has already been taken care of to define the
physical parameters, it is $\Pi^{(n)}(q_0)$ which is relevant in 
Eq.~(\ref{modpole}).

It is now simple to calculate the remaining self-energy and vertex diagrams.
A special feature is presented, however, by the constant vertex diagrams (f, g)
arising from the vertex proportional to $c_2$: While all other
contributions are proportional to $\qls$, this one is not, being given by
\be
\dab\,i\De(q)(q_\lm q_0\de_{\sg 0}+q_\sg q_0\de_{\lm 0})\,2c_2\ov n~.
\ee
It reflects the fact that our treatment breaks Lorentz invariance to $O(3)$.
Thus while the matrix element
\be
\la 0|A_\mu^a|\pi^b(q)\ra=i\de^{ab}f_\mu(q)~,
\ee
is defined in vacuum as $f_\mu=q_\mu \F$, it must be expressed in medium 
as~\cite{Pisarski,Thorsson}
\be
f_\mu=\de_{\mu 0}\,q_0\F^t+\de_{\mu i}\,q_i\F^s,~~~~~~~i=1,2,3~.
\ee

The results of calculating all the diagrams of Fig.~1 with vertices given by
(\ref{lag_int}) may now be expressed in terms of the effective parameters,
\be
m_\pi^{(n)}=m_\pi\left\{1+\left(2c_1-c_2-c_3+\frac{g_A^2}{8m_N}\right)
\frac{\ov n}{\F^2}\right\}~,
\label{effmass}
\ee
\be
\F^t=\F\left\{1+\left(c_2+c_3-\frac{g_A^2}{8m_N}\right)
\frac{\ov n}{\F^2}\right\}~,
\label{eff_t}
\ee
\be
\F^s=\F\left\{1+\left(-c_2+c_3-\frac{g_A^2}{8m_N}\right)
\frac{\ov n}{\F^2}\right\}~.
\label{eff_s}
\ee
These results were obtained earlier in this form in Ref.~\cite{Meissner}
by integrating out the nucleon field in the generating functional in 
presence of the external field $a_\mu$.
We postpone discussing the validity of these results until Sec. V.
 
\section{Virial Expansion}
\setcounter{equation}{0}
\renewcommand{\theequation}{4.\arabic{equation}}

We next turn to a different approach to the problem, namely the virial 
expansion for the self-energy of the particle in 
question~\cite{Smilga,Jeon,Mallik2}. The resulting (first order) formula 
is valid, if 
the medium is sufficiently dilute. As we shall discuss below, its range of 
validity is, in general, different from that calculated above using $\chi PT$.

Let us derive the formula for the case at hand. A simple derivation 
follows, if we recognize that the self-energy function is an $S$-matrix
element~\cite{Mallik2}. 
Consider first the process in vacuum. Just as the amplitude $T$
for the two body $\pi N$ scattering, 
\[\pi(k,i)+N(p,s)\rw\pi(\kp,\ip)+N(\pp,\spr),\] 
is given by the $S$-matrix element,
\bea
i(2\pi)^4\de^4(p+k-\pp-\kp)T_{\ip i\, ;\, \spr s} 
&=&\la\kp,\ip;\pp,\spr|S-1|k,i;p,s\ra\nonumber\\
&=&\la 0|a(\kp,\ip)b(\pp,\spr)(S-1)b^\dagger(p,s)a^\dagger(k,i)|0\ra~~,
\label{scattamp}
\eea
we may express the self-energy $\Pi^{(0)}$ of the pion (in vacuum) by the 
one-body matrix element,
\be
-i(2\pi)^4\de^4(k-\kp)\de_{\ip i}\Pi^{(0)}(k)=\la\kp,\ip|S-1|k,i\ra=
\la 0|a(\kp,\ip)(S-1)a^\dagger(k,i)|0\ra~~,
\ee
where $S$ is the familiar scattering matrix operator,
\[S=e^{i\int\,d^4x\,\cl_{int}(x)}~~.\]
Here $i(\ip)$ and $s(\spr)$ are indices denoting the pion isospin and the 
nucleon spin projection in the initial (final) state respectively. Following 
the usual practice, the
amplitude $T$ is regarded as a $2\times 2$ matrix in the nucleon isospin
space. The operators $a(k,i)$ and $b(p,s)$ annihilate respectively a pion of
momentum $k$ and isospin $i$ and a nucleon of momentum $p$ and spin
$s$.

The corresponding self-energy $\Pi (k)$ in nuclear medium is obtained 
simply by replacing the vacuum expectation value in Eq.~(4.2) by the 
ensemble average defined in Eq.~(\ref{ensavg}),
\be
-i(2\pi)^4\de^4(k-\kp)\de_{\ip i}\Pi (k)=\la a(\kp,\ip)(S-1)a^\dagger(k,i)\ra
\label{medpi}
\ee
It is here that we make use of the virial expansion in powers of the 
distribution function. We expand the ensemble average of any operator $\co$
as,
\be
\la\co\ra=\la 0|\co|0\ra+\sum_N\int\frac{d^3p}{(2\pi)^3\,2E_p}n^+(p)
\la p,s|\co|p,s\ra + \cdot\cdot\cdot~,
\ee
where the sum is over the nucleon spin and isospin states. Applying 
this expansion to the right hand side of Eq.~(\ref{medpi}), we get
\be
-i(2\pi)^4\de^4(k-\kp)\Pi^{(n)}(k)=\sum_N\int\frac{d^3p}{(2\pi)^3\,2E_p}n^+(p)
\la p,s| a(\kp,i)(S-1)a^\dagger(k,i)|p,s\ra
\ee
where $\Pi^{(n)}(k)$ stands as before for the difference, $\Pi(k)-\Pi^{(0)}(k)$.
Notice that there is no sum over the pion isospin index $i$.
We now use Eq.~(\ref{scattamp}) to express the self-energy in terms of the 
forward scattering amplitude $T_f(p,k)$,
\be
\Pi^{(n)}(k)=-\int\frac{d^3p}{(2\pi)^3\,2E_p}n^+(p)\sum_N T_f(p,k)~,
\label{virial}
\ee
taking the summation inside the integral, as the distribution function is 
the same for all the four nucleon states. 

Eq.~(\ref{virial}) is the desired 
first order virial expansion formula for the pion self-energy. Although such
formulae have been used to find the mass shifts in different cases, 
its application to pion in nuclear medium does not exist in 
the literature. What has been utilised earlier is an approximation to the 
above formula~\cite{Migdal}
\be
\Pi^{(n)}(k)=-\frac{\ov n}{8m_N}\sum_N\,T_f(k)~.
\ee
But if the amplitude is not constant, even approximately, within the
interval of integration in Eq.(4.6), this formula cannot clearly be trusted. As we 
shall see below, this is indeed the situation for the case at hand.

To sum over the forward amplitudes, we begin with the kinematics
for the process in general. The matrix structure of $T$ in nucleon isospin
space may be written as
\be
T_{\ip i\,;\,\spr s}=
T^+_{\spr s}\de_{\ip i}+T^-_{\spr s}\frac{1}{2}[\tau_{\ip},\tau_i],
\ee
where each of the $T^\pm$ has the invariant spin decomposition,
\be
T^\pm_{\spr s}=
\bu(\pp,\spr)\{A^\pm+\frac{1}{2}(\kp\!\!\!\!\!/+k\!\!\!/)B^\pm\}u(p,s)~.
\ee
Taking the amplitude for any one of the charged states $\pi^{\pm,0}$
for the pion, we can readily carry out the sum over the nucleon states,
\be
\sum_N T_f(p,k)=8(m_NA^++p\cdot kB^+)~,
\ee
in terms of the isospin even amplitude only.
							
We now evaluate $\Pi^{(n)}$ using the experimental data for the $\pi N$ scattering
amplitude~\cite{Hohler}. As given by Eq.(4.6), it is an integral over the 
3-momentum of the nucleon in the pion rest frame $(\vk=0)$, while the data is
given as a function of the pion energy $\omk=\sqrt{\vk^2+m_\pi^2}$
in the nucleon rest (lab) frame. The two variables are related by the equation,
$m_\pi E_p=m_N\omk$. We thus have finally the complex pole position in the 
pion propagator as
\be
m_\pi^{(n)}-\frac{i}{2}\gamma_\pi^{(n)}=
m_\pi+\frac{\Pi^{(n)}}{2m_\pi}=
m_\pi-\frac{1}{\pi^2}\left(\frac{m_N}{m_\pi}\right)^3
\int_1^{\ov\om_k}\,d\omk\sqrt{\omk^2-m_\pi^2}\,D^+(\omk)
\ee
where $D^+(\omk)= A^++\omk B^+$, is the isospin even forward
$\pi N$ scattering amplitude. The upper limit $\ov\om_k$ is determined
by the nucleon number density
\be
\ov\om_k=m_\pi \sqrt{1+\left ( \frac{3\pi^2\ov n}{2m_N^3}\right )^{2/3}}~.
\ee
The imaginary part of the pole position represents the damping rate of 
pionic excitations in the medium.

The numerical evaluations are shown in Figs. 2 and 3, where the pion mass
shift and its imaginary part are plotted as a function of the nucleon number
density in units of the normal density ${\ov n}_0$=(110 MeV)$^3$.
Our result for the pion mass shift in nuclear medium may be compared with 
that for the nucleon mass shift in pionic medium, calculated in~\cite{Smilga}.
It will be observed that while both the shifts are given essentially by
integrals over the same $\pi N$ amplitude times the corresponding distribution
functions, the curves bend in opposite directions. The reason is that as the
pion energy increases, there is a change in sign in the real part of the 
amplitude, which is weighted differently by the distribution functions
in the two cases.

\bef[h]
\centerline{\psfig{figure=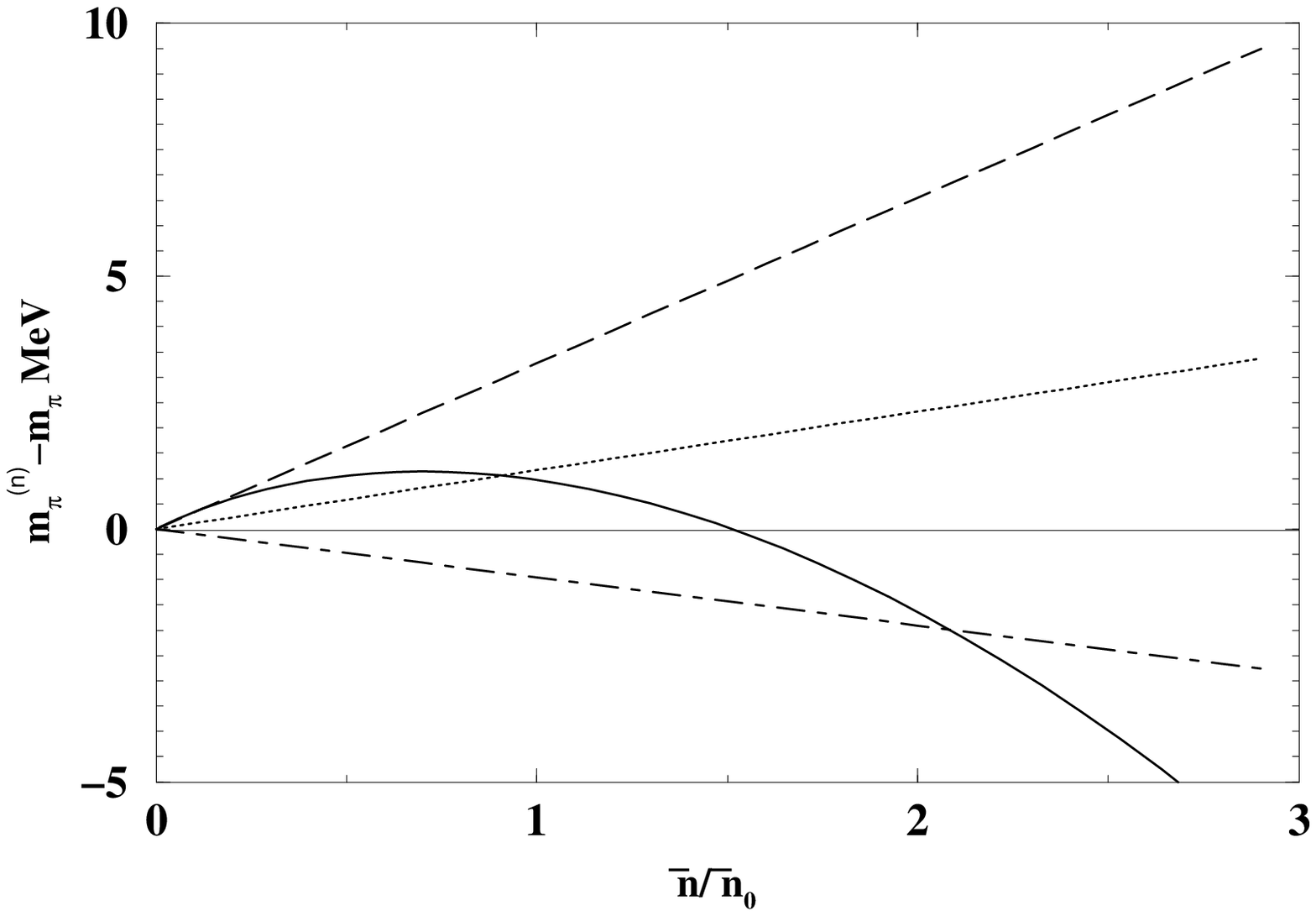,height=7cm,width=7.5cm}}
\caption{Shift in pion mass in nuclear medium. The solid curve results from 
the virial formula. The three straight lines follow from $\chi PT$ : the
dotted, the dashed and the dash-dotted curves correspond to the central
value, the upper and the lower limits of the constants in Eq.(\ref{effmass}).} 
\eef

\bef[h]
\centerline{\psfig{figure=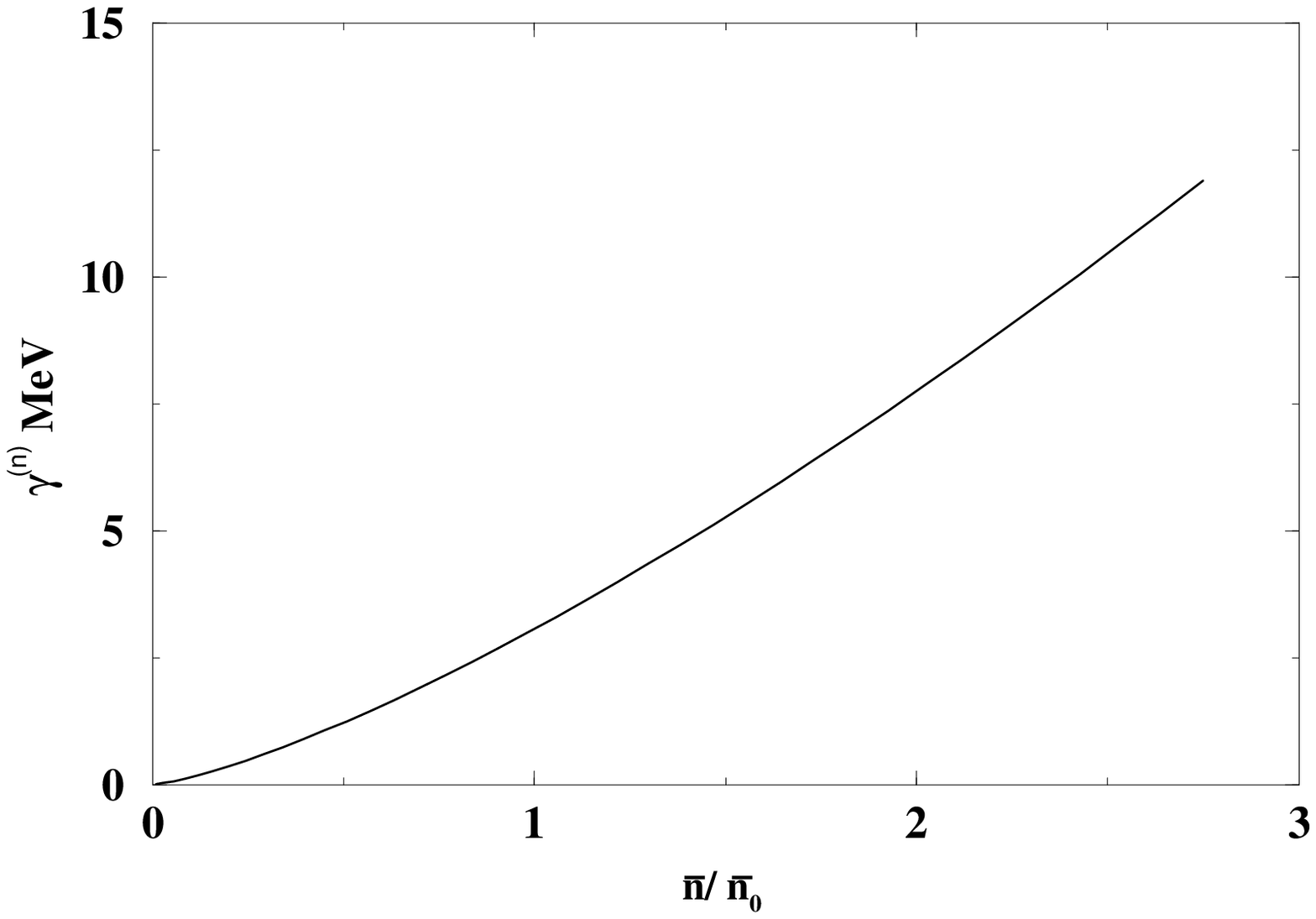,height=7cm,width=7.5cm}}
\caption{Damping rate of pionic excitations in nuclear medium.}
\eef

\section{Discussion}

Having derived the effective parameters of the pion in nuclear medium by two
different methods, we now discuss the validity of the results. Considering 
$\chi PT$,
the region in which the leading correction term may represent a meaningful 
approximation to a quantity depends on the proximity of the resonances in the
relevant channel. In the present case, a number of resonances, particularly the
$\Delta (1232)$, lie close to the threshold, making the values of the
coupling constants $c_1,\,c_2$ and $c_3$ in the effective Lagrangian rather 
large. Then the region of nuclear density in
which the results may be valid, is expected to shrink considerably. 

The calculated results follow this expectation. In the expression 
(\ref{effmass})
for the effective mass, there is, however, a large cancellation among the
contributions of the different vertices, making the mass shift to be only a few
MeV at $\ov n=\ov n_0$, the normal nuclear density. But in the 
expression (\ref{eff_s}) for $\F^{s}$ the contributions of the vertices add up,
making the "correction" at this density overwhelm the unit term. Clearly, 
the first order $\chi PT$ results for the 
pion traversing nuclear matter at normal density are unacceptable.

On the other hand, in the virial expansion formula we may avoid any
inaccuracy in calculating the scattering amplitude by taking it from
experiment. However, the nuclear medium must be dilute enough for an
expansion of the self-energy function in powers of nuclear density to be 
valid. At normal nuclear density $\ov n_0=(110 \,$MeV)$^3$, the mean distance 
between the nucleons is about 2 fm. 
Thus our first order virial formula, where the pion 
propagation is perturbed by a single interaction with one of the nucleons in 
the medium, should be a reasonable approximation up to about this density.
We point out that we
use the exact formula, rather than the approximate version used so long
by different authors in this context. Unfortunately, the virial expansion 
does not say anything about the residues, $\F^t $ and $\F^s$. At normal
nuclear density, its prediction
for the mass-shift and the decay width of the pion is that both are
negligible, being only a few MeV.  

We conclude with a comment on determining the mass shift in the more 
interesting case of the nucleon in nuclear medium. Here the simple chiral 
Lagrangian
for the $NN$ system~\cite{Weinberg2} is not expected to apply, as it does not 
take properly into account the presence of bound or virtual two nucleon 
states close to the $NN$ threshold. In fact, it produces an absurdly large 
value for the nucleon mass shift~\cite{Montano}. But the virial expansion 
formula for the nucleon self-energy should apply here, at least in the 
neighbourhood of the nuclear saturation density~\cite{xyz}.

\section*{acknowledgements}

We are grateful to J. Gasser and H. Leutwyler for correspondence. We thank
the Referee for his criticism and referring us to the relevant literature. 
One of us (S.M.) acknowledges the support of CSIR, Government of India.

\end{document}